\documentclass[twocolumn,aps,superscriptaddress]{revtex4}
\usepackage{amssymb}
\usepackage{amsmath,bm}
\usepackage{graphicx}
\usepackage[normalem]{ulem}

\usepackage[usenames, dvipsnames]{xcolor}
\usepackage{lipsum}

\renewcommand{\sout}{\bgroup \color{red} \ULdepth=-.5ex \ULset}

\begin{document}
\title{Pairing effects on neutron matter equation of state and symmetry energy
at subsaturation densities}
\author{Zhen Zhang\footnote{zhangzh275@mail.sysu.edu.cn}}
\affiliation{Sino-French Institute of Nuclear Engineering and Technology, Sun Yat-Sen University,
Zhuhai 519082, China}
\author{Lie-Wen Chen\footnote{lwchen$@$sjtu.edu.cn}}
\affiliation{School of Physics and Astronomy and Shanghai Key Laboratory for
Particle Physics and Cosmology, Shanghai Jiao Tong University, Shanghai 200240, China}
\date{\today}

\begin{abstract}
Within the framework of BCS theory and Skyrme-Hartree-Fock model, we employ various microscopic
pairing gaps and effective
pairing interactions to study  pairing effects on
the equation of state (EOS) of neutron matter and the symmetry energy at subsaturation densities. We find
pairing effects may have considerable contributions to the EOS of neutron matter at very low densities ($\lesssim 0.02~\rm{fm}^{-3}$),
while only have a small impact on the symmetry energy at subsatruation densities.
In addition, the reliability
of the parabolic approximation for the isospin asymmetry dependence of nuclear matter EOS 
with pairing correlations included is also discussed.

\end{abstract}


\maketitle
\section{Introduction}
Nuclear symmetry energy $E_{\mathrm{sym}}(\rho)$ and the equation of state (EOS)
of pure neutron matter (PNM) $E_{\mathrm{PNM}}(\rho)$ have profound impact on many important
physics problems in nuclear physics and astrophysics\cite{Lat04,Ste05,Bar05,LCK08,Oer17} as well
as some issues
in new physics beyond the standard model~\cite{Hor01,Sil05,WLC09,ZZC14,ZSC15}. 
For instance, the density dependence of the symmetry energy
or neutron matter EOS at subsaturation densities is intimately related
to the neutron skin thickness 
of finite nuclei~\cite{Bro00,Bro13,Typ01,Hor01a, Fur02, Yos04, Tod05,CLK05,CKLX10,Cen09,Roc11,Agr12,Zha13}, 
the properties of neutron star crust~\cite{Hor01a, Xu09, Duc11,Zha14},
the cluster formations in nuclear matter at low densities~\cite{Kow07,Nat10,Wad12,Typ10,ZhangChen17}, 
and the isospin diffusion in  heavy ion
collisions at Fermi energies~\cite{Tsa04,Tsa09,CKL05,LC05}.

In last few decades, a lot of efforts based on various phenomenological models 
or microscopic theories have been devoted to exploring
$E_{\mathrm{sym}}(\rho)$ and $E_{\mathrm{PNM}}(\rho)$ at subsaturation densities.
From analyses of various experimental observables
using phenomenological models, our knowledge on nuclear matter EOS 
at subsaturation densities have been significantly improved. 
For example, in the density region of about $\rho_0/3\sim\rho_0$ 
($\rho_0\approx0.16~\mathrm{fm}^{-3}$ is the nuclear saturation density), 
constraints on nuclear symmetry energies have been extracted from 
analyses of nuclear masses~\cite{Zha13,Bro13} as well as isobaric analog states and 
neutron skin data~\cite{Dan14} using Skyrme-Hartree-Fock (SHF) model,
and also from transport model analyses of experimental observables in
mid-peripheral heavy ion collisions of Sn isotopes~\cite{Tsa09};
around $\rho_0/3$, both $E_{\mathrm{sym}}(\rho)$ and $E_{\mathrm{PNM}}(\rho)$ 
have been well determined by the electric dipole
polarizability in $^{208}$Pb~\cite{Zha15}. 
These constraints from phenomenological models are essentially
consistent and the uncertainty can be as small as $\approx 1~\mathrm{MeV}$ (see, e.g., Ref.~\cite{Zha15}). 

In addition, great progress in constraining $E_{\mathrm{sym}}(\rho)$ or 
$E_{\mathrm{PNM}}(\rho)$ at subsaturation densities has also been made by ab initio 
calculations, especially with the development of modern chiral nuclear force. 
For examples, the QCD sum rules have been shown to provide important 
information on EOS of neutron matter at subsaturation densities~\cite{Cai18}.
Based on modern chiral nuclear forces, various many-body theories
such as chiral effective field theory~\cite{Mac11,Tew13}, 
self-consistent Green function method~\cite{Car14}, coupled-cluster 
method~\cite{Hag14} and Quantum Monte Carlo technique~\cite{Car15} 
have also provide important information from first principle 
on nuclear matter EOS at subsaturation densities.

It should be pointed out that the constraints
on $E_{\mathrm{sym}}(\rho)$ or $E_{\mathrm{PNM}}(\rho)$ from
phenomenological models (e.g., SHF model
and relativistic mean field theory) are usually obtained
without including nucleon pairing correlation, while many ab initio calculations
include it.  
It is well known
that the $^1\mathrm{S}_0$ pairing gap in nuclear matter can
play an essential role at low densities (e.g., $\rho_0/10$ )~\cite{Lom99,Dea03}.
Therefore,  it is important and interesting to examine the pairing
effects on the EOS of nuclear matter at subsaturation densities.
This is especially the case when the phenomenological models give more and more 
stringent constraints on the $E_{\mathrm{sym}}(\rho)$ and $E_{\mathrm{PNM}}$ 
at subsaturation densities (see, e.g., Ref.~\cite{Zha15}), 
which provides the main motivation of the present work.

Theoretically great efforts have been made to determine 
the density dependence of pairing gap in neutron matter using various 
microscopic many-body approaches, e.g., 
Bardeen-Cooper-Schrieffer~(BCS) theory~\cite{Heb07, Zha10}, 
correlated-basis-function calculations~\cite{Che93, Fab05}, 
renormalization group~\cite{Sch03}, Brueckner theory~\cite{Cao06} and 
Quantum Monte Carlo~\cite{Gez10,Gan08,Gan09}. 
In particular, at extremely low densities, or equivalently low 
$k_{\mathrm{F}}a$ with $a\approx-18.5$ fm being neutron scattering length
and $k_{\mathrm{F}}$  neutron Fermi momentum, 
there exists a well-known analytical pairing gap
of $\Delta(k_F)=\frac{1}{(4e)^{1/3}}\frac{8}{e^2}\frac{\hbar^2k_F^2}{2M}\mathrm{exp}\left( \frac{\pi}{2ak_F}\right)$ 
with $e$ being Euler's number~\cite{Gor61}. 
Thanks to the analytical limit and microscopic calculations, 
the pairing gap in neutron matter at densities $< \rho_0/10$ 
is under good control. 
But its higher density behaviors are still largely uncertain~\cite{Gez14}. 
Moreover, 
the isospin-dependence of the pairing gap is even more poorly known
and microscopic calculations are still inadequate for the study of 
pairing effects on  the symmetry energy. 
An alternative perspective is starting from effective pairing interactions
which are usually constructed by fitting properties of finite
nuclei~{\cite{Kha09,Yam09, Yam12} or microscopic pairing gaps~\cite{Mar07} 
with a hypothetical functional form. Using the effective pairing interactions,
one can study the effects of pairing correlation on nuclear symmetry energy 
within the framework of BCS theory.

In this work, within the framework of BCS theory together with the SHF model, 
we study pairing effects on the EOS of neutron matter at subsaturation densities by invoking various 
microscopic pairing gaps or effective pairing interactions.
Subsequently, we use several effective pairing interactions to study pairing 
effects on the symmetry energy at subsaturation densities. 
The reliability of the parabolic approximation for the isospin 
asymmetry dependence of nuclear matter EOS is further verified in the case of including 
the effect of pairing correlations.

The paper is organized as follows. 
In Sec.~\ref{Sec:PF} we introduce the effective pairing interactions used in this work 
and the calculation of pairing energy density. 
Sec.~\ref{Sec:Results} presents the results for pairing effects on neutron matter EOS 
and the symmetry energy at subsaturation densities. 
We then end the paper with a summary and conclusions.

\section{Effective pairing interaction and pairing energy density}\label{Sec:PF}
In the framework of  BCS theory, the equation for the pairing gap 
$\Delta_q$ ($q=n, p$) in nuclear matter is given by
\begin{equation}
\label{Eq:gap}
\Delta_q(k)=-\int \frac{d^3k'}{(2\pi)^3} \frac{v_q(k,k')\Delta_q(k')}{2\sqrt{[\epsilon_q(k')-\lambda_q]^2+\Delta_q^2}},
\end{equation}
where $v_q$ is the pairing strength in momentum space, 
$\lambda_q$ is the effective chemical potential, 
and $\epsilon_q(k)=\hbar^2k^2/2m_q^{\ast}$ is the single-particle 
kinetic energy with $m_q^*$ being the nucleon effective mass.
It should be pointed out that in the case of contact pairing 
interaction, both $v_q$ and $\Delta_q$ 
are momentum independent. 
Note that the momentum independent mean-field potentials have been 
absorbed into the effective chemical potential, which can be determined 
by the nucleon density
\begin{equation}
\label{Eq:den}
\rho_q = \int \frac{d^3k}{(2\pi)^3}\left[ 1-\frac{\epsilon_q(k)-\lambda_q}{E_q}\right],
\end{equation}
where $E_q = \sqrt{\left[\epsilon_q(k)-\lambda_q\right]^2+\Delta_q^2}$ 
is the quasi-particle energy.
For contact pairing interactions,
once given the pairing strength $v_q(\rho_n, \rho_p)$ [the pairing gap $\Delta_q(\rho_n, \rho_p)$] and the nucleon effective 
mass $m_q^{\ast}(\rho_n, \rho_p)$ at neutron density $\rho_n$ and proton density 
$\rho_p$, the $\Delta_q(\rho_n, \rho_p)$  [$v_q(\rho_n, \rho_p)$] and effective chemical 
potential $\lambda_q(\rho_n, \rho_p)$ can be easily obtained by solving 
Eqs.~(\ref{Eq:gap}) and~(\ref{Eq:den}).
The pairing energy density in asymmetric nuclear matter 
then can be expressed as
\begin{eqnarray}
\label{Eq:enden}
\varepsilon_{\rm{pair}}&=&\sum_{q=n,p}\int \frac{d^3k}{(2\pi)^3}
\left\lbrace \epsilon_q(k)\left[1-\frac{\epsilon_q(k)-\lambda_q}{E_q} \right]
-\frac{1}{2}\frac{\Delta_q^2}{E_q} \right\rbrace  \notag\\
&&-\sum_{q=n,p} \frac{3}{5}\frac{\hbar^2}{2m_q^{\ast}}\rho_q k_{F,q}^2 ,
\end{eqnarray}
where $k_{F,q}=(3\pi^2\rho_q)^{1/3}$ is the Fermi momentum. 
In the weak coupling approximation 
($\Delta_q\ll \hbar^2k^2_{F,q}/2m_q^{\ast}$)~\cite{Cha10}, 
the pairing energy density can be approximated as
\begin{equation}
\label{Eq:ana}
\varepsilon_{\rm{pair}}=-\frac{1}{2}\left( N_n\Delta_n^2+N_p\Delta_p^2\right),
\end{equation}
with $N_q = m_q^*k_{F,q}/2\pi^2\hbar^2$ being the density of states.

While the pairing effect on neutron matter EOS can be studied
by directly invoking microscopic pairing gaps,
for the symmetry energy, due to our poor knowledge on the 
isospin dependence of pairing gaps, one has to introduce effective pairing 
interactions.
In this work, following Ref.~\cite{Kha10,Mar14}, we use several 
different effective contact pairing interactions to study the 
pairing effects on nuclear symmetry energy at subsaturation
densities.
Note that the integral in Eq.(\ref{Eq:gap}) is divergent for contact 
pairing interactions.
Therefore, cutoff momenta are usually introduced in the 
effective contact pairing interactions in different prescriptions 
corresponding to different physics problems
(see, e.g., Ref.~\cite{Mar07}). In the construction of pairing interactions
used in this work, the cutoff is defined with respect
to the quasiparticle energy $\sqrt{(\epsilon_q(k)-\lambda_q)^2
+\Delta_q^2}<E_C$, where $E_C$ is the cutoff energy and is taken to 
be $60$ MeV~\cite{Yam12,Mar07}.

The commonly used effective contact pairing interaction has the form of
\begin{equation}
v_q(\bm{r},\bm{r}^{\prime})
=v_0\left[1-\eta\left( \frac{\rho}{\rho_c}\right)^{\alpha} \right]
\delta(\bm{r}-\bm{r}^{\prime}),
\end{equation}
with $\rho_c =0.16~\rm{fm}^{-3}$ and $v_0$ being the strength parameter. 
The parameter $\alpha$ is usually taken to be one for simplicity, and then 
the parameter $\eta$ determines the density dependence of the pairing 
interaction with $\eta=1$ for surface pairing and $\eta=0$ for volume pairing. 
In recent studies, $\eta= 0.5$ (mix pairing) is preferred as
it can well reproduce the mass dependence of the odd-even mass staggering 
parameter~\cite{Dob02}.

Considering the isospin dependence of the pairing gaps, an extended
pairing interaction with the inclusion of isovector density $\rho_1=\rho_n-\rho_p$
has been introduced in Ref.~\cite{Yam09, Yam12} as
\begin{eqnarray}
v_q(\bm{r},\bm{r}^{\prime})&=&
v_0\left[ 1-\eta_0\frac{\rho}{\rho_0}-\eta_1\tau_3\frac{\rho_1}{\rho_0}-\eta_2\left(\frac{\rho_1}{\rho_0} \right)^2 \right] \notag \\
& &\times \delta(\bm{r}-\bm{r}^{\prime}),
\end{eqnarray}
where $\tau_3 = 1(-1)$ for $q=n(p)$.
The additional $\rho_1$ terms are important to describe the isospin dependence 
of experimental pairing gaps. 
In this work, we consider three parametrizations from Ref.~\cite{Yam12}:
1) $v_0 = -370.8~\mathrm{MeV\ fm}^3$, $\eta_0=0.75$ and $\eta_1=\eta_2=0$ (SLy4+IS);
2) $v_0 = -396.47~\rm{MeV\ fm}^3$, $\eta_0=0.75 $, $\eta_1 = 0.270$ and $\eta_2=2.5$ (SLy4+IV); 3)$v_0 = -388.60~\rm{MeV\ fm}^3$, $\eta_0=0.75 $, $\eta_1 = 0.4$ and $\eta_2=2.5$ (LNS+IV}). 
Here we denote them as `IS' and `IV' since they are isospin-independent 
(scalar) and isospin-dependent (vector), respectively.
All the three parametrization forms 
are optimized to fit experimental pairing gaps in Hartree-Fock-Bogoliubov 
calculations using the SLy4~\cite{Cha98} or the LNS~\cite{Cao06a} Skyrme interaction. 
It should be noticed that the SLy4 predicts negative neutron-proton effective mass splitting, i.e., $m_n^*<m_p^*$, in neutron-rich matter, while the LNS interaction, which is obtained by fitting predictions of Brueckner-Hartree-Fock calculations, has $m_n^*>m_p^*$. As can be seen from Eq.~(\ref{Eq:ana}), the isospin-dependence of the nucleon effective mass is related to the isospin behavior of the pairing energy. 
We will also consider the isospin dependent SLy4+MSH~\cite{Mar07}  pairing interaction 
which is given as~\cite{Mar07}
\begin{eqnarray}
v_q^{\mathrm{MSH}}(\bm{r},\bm{r}^{\prime})=v_0
\left[ 1-(1-\tau_3\delta)\eta_s\left( \frac{\rho}{\rho_0} \right)^{\alpha_s}\right. \notag \\
\left.-\tau_3\delta\eta_n\left( \frac{\rho}{\rho_0} \right)^{\alpha_n}\right]
\delta(\bm{r}-\bm{r}^{\prime}),
\end{eqnarray}
with $v_0=-448~\mathrm{MeV~fm}^{3}$, $\eta_s=0.598$, $\alpha_s=0.551$,
$\eta_n=0.947$, $\alpha_n=0.554$,  and $\delta=(\rho_n-\rho_p)/\rho$ 
being isospin asymmetry. 
The SLy4+MSH parameter set
is determined by fitting neutron pairing gaps in both symmetric 
nuclear matter and pure neutron matter predicted by microscopic Brueckner 
calculations~\cite{Cao06} with the effective mass deduced from 
the SLy4 interaction.

\section{Results and Discussion}~\label{Sec:Results}

\begin{figure}[!hpbt]
\includegraphics[width=1.\linewidth]{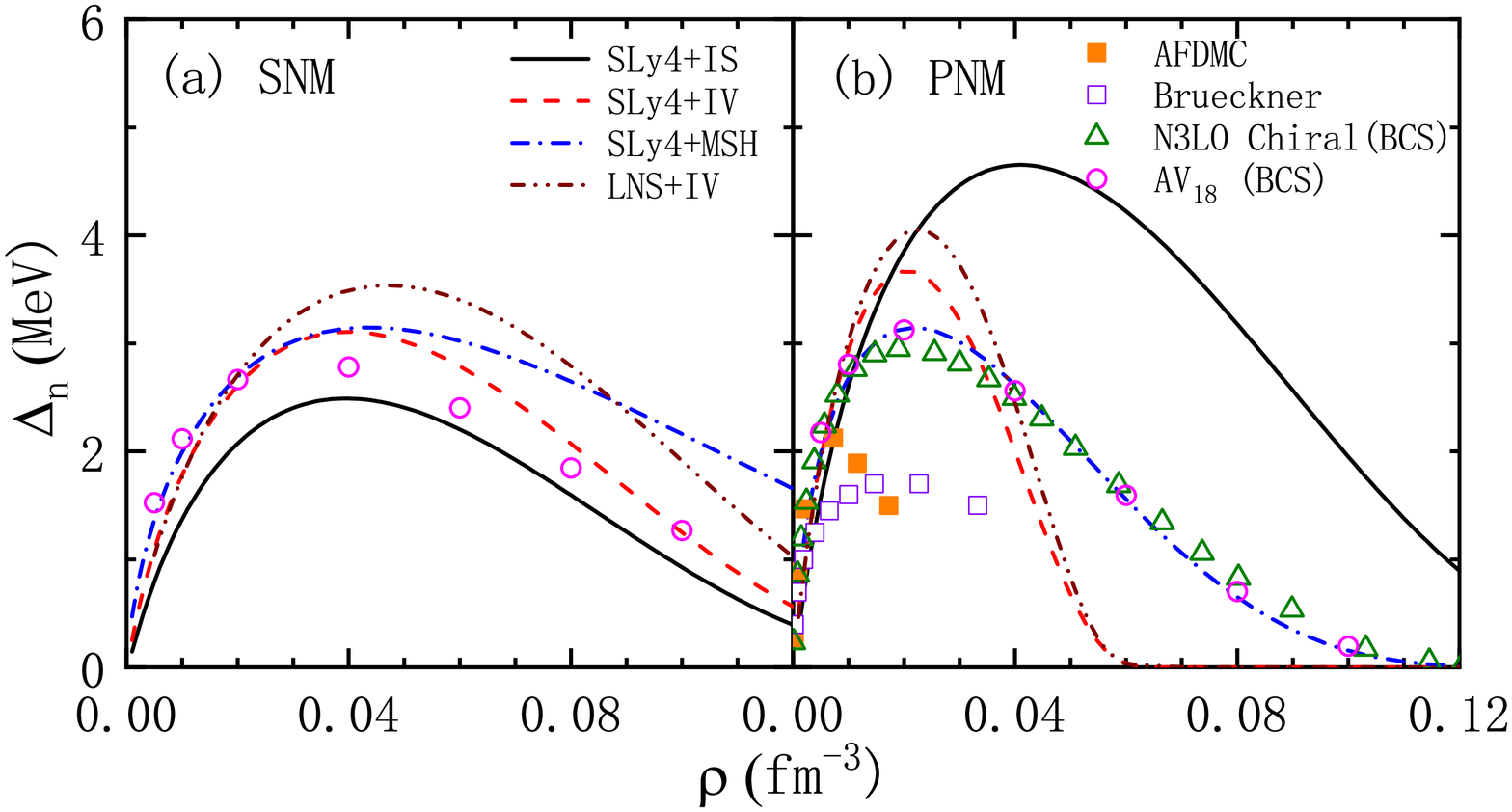}
\caption{Density dependence of
$^1$S$_0$ neutron pairing gap in symmetric nuclear matter (a)
and pure neutron matter (b) from BCS calculations using various
effective pairing interactions and using the microscopic Argonne V18 two-body force with the LNS mean-field~\cite{Zha10}.
Panel (b) also includes corresponding results from calculations using microscopic nuclear forces (see the text for details).}
\label{Fig:gap}
\end{figure}

In Fig.~\ref{Fig:gap}, we show the density dependence
of neutron $^1$S$_0$ pairing gaps in symmetric nuclear matter 
(SNM) (panel (a)) and PNM (panel (b)) 
obtained from the BCS calculations using SLy4+IS, SLy4+IV, SLy4+MSH and LNS+IV, as well as using the microscopic Argonne V18 two-body force~\cite{Zha10}  (denoted as AV$_{18}$ (BCS)).
Also included in Fig.~\ref{Fig:gap}~(b) are the corresponding results from
three different types of pairing gaps in pure neutron matter predicted by different methods:
`N3LO Chiral (BCS)' is calculated using chiral N3LO two-body interaction at
BCS level~\cite{Heb07}; `Brueckner' is the prediction
of Brueckner theory using the Argonne V18 two-body force and a three-body 
force~\cite{Cao06};
`AFDMC' is obtained from the auxiliary field diffusion
Monte Carlo (AFDMC) method calculation~\cite{Gan08}
using  realistic  two- and three-body nuclear force, i.e., the Argonne $v'_8$ 
and the Urbana IX.
It is seen from Fig.~\ref{Fig:gap} that pairing gaps from the various effective pairing interactions or approaches have quite different density dependence, especially in neutron matter.
Note that the pairing gaps in neutron matter from the BCS calculations with 
realistic two-body interactions [i.e.,N3LO Chiral (BCS) and AV$_{18}$ (BCS) ]   are in very good agreement with 
each other and roughly reflect the upper limit of microscopic
gaps (see e.g. Ref.~\cite{Gez14}) at low densities. 
As shown in Fig.\ref{Fig:gap}~(b), the pairing gaps in PNM from 
the effective contact pairing interactions and microscopic nuclear 
forces are rather different, especially at higher densities.

\begin{figure}[!pbt]
\includegraphics[width=\linewidth]{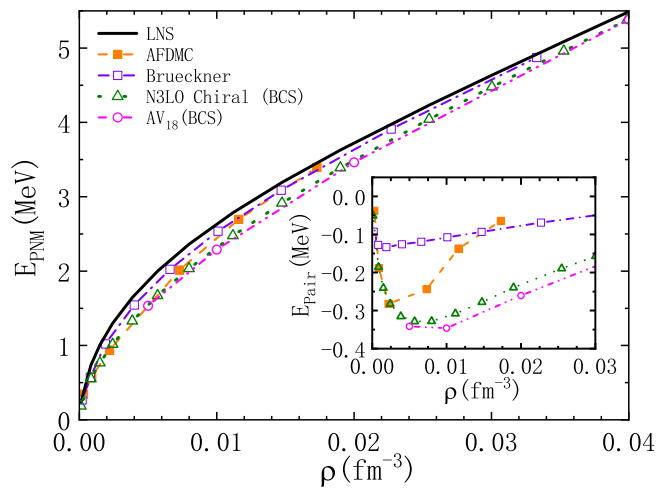}
\caption{
Binding energy per neutron in pure neutron matter as a function of density
without and with different density-dependent pairing gaps (see the text for details).
The inset shows the difference between the neutron matter EOSs with and
without pairing correlation.}
\label{Fig:Epnm}
\end{figure}

\begin{figure*}
\includegraphics[width = \linewidth]{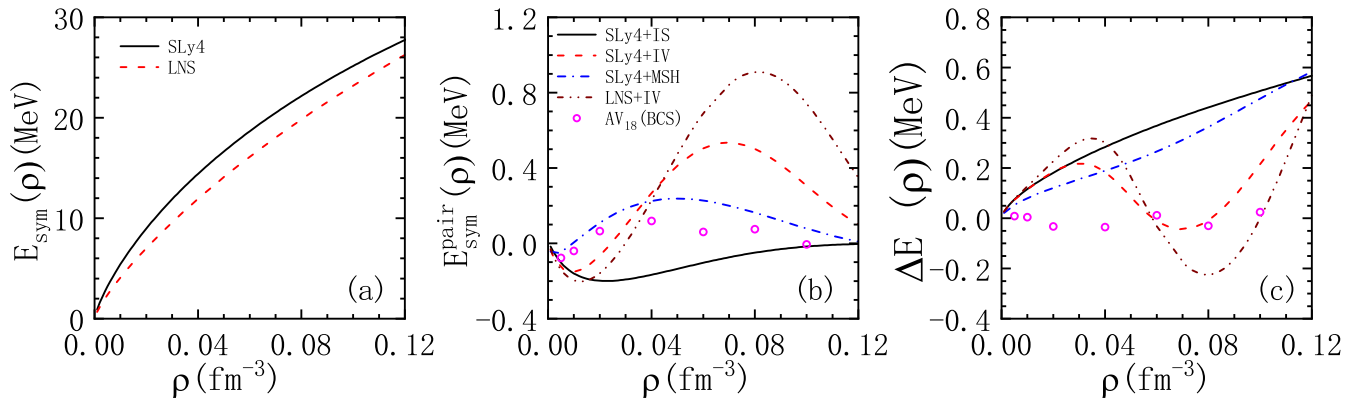}
\caption{(a) Symmetry energy as functions of density calculated using SLy4 and LNS
interactions without pairing correlations.
(b) Contributions of pairing effects on the symmetry energy
(c) Deviations of the symmetry energy in parabolic approximation from
its exact value calculated using different pairing interactions.}
\label{Fig:Esym}
\end{figure*}

As introduced in Sec.~\ref{Sec:PF}, for contact pairing interaction, the pairing strength can be exactly determined by the pairing gap and the nucleon effective masse. Therefore, one can directly construct contact pairing interactions from microscopic pairing gaps and further study pairing effects on properties of nuclear matter and finite nuclei~\cite{Cha10,Gor09}.  
Combining the LNS Skyrme interaction with the four neutron pairing gaps 
in PNM from microscopic nuclear interactions, we calculate the EOS of 
pure neutron matter at low densities up to $0.04~\mathrm{fm}^{-3}$ and
the results are shown in Fig.~\ref{Fig:Epnm}. For comparison, Fig.~\ref{Fig:Epnm} also includes the prediction 
of Hartree-Fock calculation using LNS interaction without 
neutron pairing correlation.
It is seen that the neutron pairing has negative contribution to the 
EOS of PNM, especially around the density of $ \rho =0.005~\mathrm{fm}^{-3}$.
To more clearly clarify the effect of pairing correlation, we define 
$E_{\mathrm{pair}}$ as the difference between the EOSs
with and without neutron pairing, and exhibit $E_{\mathrm{pair}}$ as a 
function of neutron density in the inset of Fig.~\ref{Fig:Epnm}. 
One sees that the neutron pairing may play a considerable role at 
very low densities (e.g., in the case of `N3LO Chiral (BCS)' and `AV$_{18}$(BCS)', 
the neutron matter EOS is reduced by about $20\%$ at densities around 
$\rho= 0.005~\mathrm{fm}^{-3}$), but its effect turns to be negligible 
towards higher densities (e.g., $E_{\mathrm{pair}}<0.25~\mathrm{MeV}$ 
above about $\rho= 0.02~\mathrm{fm}^{-3}$).
We would like to point out that the conclusion is essentially
independent on the choice of Skyrme interaction,
since at such low densities the effective masses of neutron are
approximately equal to the neutron bare mass.

Furthermore, we study the pairing effects on the symmetry energy 
using the isospin-dependent pairing gap `AV$_{18}$ (BCS)'~\cite{Zha10} and the four effective pairing interactions, i.e., SLy4+IS, SLy4+IV, LNS+IV and SLy4+MSH 
interactions introduced in Sec.~\ref{Sec:PF}.
In Fig.~\ref{Fig:Esym}~(a), we show the density dependence of symmetry energy
at subsaturation densities  below $0.12~\mathrm{fm}^{-3}$ obtained from Hartree-Fock calculations
using SLy4 and LNS interactions without including pairing correlation. 
Fig.~\ref{Fig:Esym}~(b) shows contributions of pairing correlation 
on the symmetry energy, $E_{\mathrm{sym}}^{\mathrm{pair}}(\rho)$, 
for the `AV$_{18}$ (BCS)' and the four effective pairing interactions.
Here $E_{\mathrm{sym}}(\rho)$ is numerically calculated according to its 
definition $E_{\mathrm{sym}}(\rho)=\left.\frac{1}{2!}\frac{\partial ^2E(\rho,\delta)}{\partial \delta^2}\right\vert_{\delta=0}$ 
with $E(\rho,\delta)$ being the EOS of 
asymmetric nuclear matter.
It can be seen that the `AV$_{18}$ (BCS)' pairing gaps lead to rather small and thus negligible  effects on the symmetry energy. For effective pairing interactions, while the SLy4+IS interaction always reduces
the symmetry energy, SLy4+MSH, SLy4+IV and LNS+IV interactions can provide either 
positive or negative contributions depending on the density.
For the three effective pairing interactions combined with SLy4 interaction, 
the magnitude of $E_{\mathrm{sym}}^{\mathrm{pair}}(\rho)$ is less than 
about $0.5$~MeV, while in the case of LNS+IV, the $E_{\mathrm{sym}}^{\mathrm{pair}}(\rho)$ reaches a maximum of about  $0.9$ MeV at the density of $\rho=0.08~\mathrm{fm}^{-3}$. Although the SLy4+IV and LNS+IV pairing interactions are constructed by fitting the same experimental data with the same form of parametrization, the effect of LNS+IV on the symmetry energy is much larger than that of SLy4+IV, which could be related to their different isospin dependence of nucleon effective masses, i.e.,the SLy4 and LNS interactions predict respectively negative and postitive neutron-proton effective mass splittings  in neutron-rich nuclear matter. 
It is seen that the maximum of $E_{\mathrm{sym}}^{\mathrm{pair}}(\rho)$  is only about $4.5\% $ of the $E_{\mathrm{sym}}(\rho)$ (about $20$ MeV) 
and thus usually can be neglected. 
Nevertheless, with the more in-depth research and more precise constraints 
on the symmetry energy at low densities, 
the pairing effects would become nonnegligible.
At this point, we
would also like to point out that the cluster formations in 
nuclear matter at subsaturation densities  have a substantial 
impact on nuclear symmetry energy~\cite{Nat10,Typ10,ZhangChen17}. 
It would be nice to investigate simultaneously the effects of pairing
correlations and cluster ormation in a self-consistent framework, 
and this may be pursued in future.

It is also interesting to check the pairing effects on the applicability of the widely 
used parabolic approximation for the isospin asymmetry dependence of nuclear matter EOS, 
in which the symmetry energy is approximated to be
$E_{\mathrm{sym}}^{\mathrm{PA}}=E_{\mathrm{PNM}}(\rho)-E_0(\rho)$.
To this end, we present in Fig.~\ref{Fig:Esym}(c) the deviation of 
the symmetry energy in parabolic approximation from its exact value, namely, 
$\Delta E(\rho)=E_{\mathrm{sym}}^{\mathrm{PA}}(\rho)- E_{\mathrm{sym}}(\rho)$, 
for different pairing interactions. 
It is seen that the $\Delta E(\rho)$ is negligible compared with
the magnitude of $E_{\mathrm{sym}}(\rho)$ (the relative deviation is about $2\%$).
One can thus conclude that the parabolic approximation 
for the isospin asymmetry dependence of nuclear matter EOS is still valid for the EOS of asymmetric nuclear matter with contributions from pairing correlations.

\section{Conclusion}
Within the framework of BCS theory together with the SHF model, 
we have investigated the pairing effects on the equation of
state of pure neutron matter and the symmetry energy at
subsaturation densities.
For neutron matter, 
invoking microscopic pairing gaps together with the LNS Skyme interaction, 
we have found  the pairing correlations may have essential impact on 
the EOS of neutron matter at very low densities below about $0.02~\mathrm{fm}^{-3}$, 
while they turn to be negligible at higher densities.
For symmetry energy, using various effective pairing
interactions and the `AV$_{18}$ (BCS)', we have found the inclusion
of pairing correlations only slightly affect the magnitude of the
symmetry energy and usually can be safely neglected. 
In addition, 
the parabolic approximation for the isospin asymmetry dependence of 
nuclear matter EOS has been proved to be a reasonable
approximation in the case of including the pairing correlations.

\begin{acknowledgments}
We thank Shi-Sheng Zhang for providing us the isospin-dependent pairing gap from BCS calculation using the Argone V$_{18}$ two-body nuclear force.
This work was supported in part by the National Natural Science Foundation of 
China under Grant No. 11625521 and 11905302, the Major State Basic Research Development 
Program (973 Program) in China under Contract No. 2015CB856904, 
the Program for Professor of Special Appointment (Eastern Scholar) at 
Shanghai Institutions of Higher Learning, Key Laboratory for Particle Physics, 
Astrophysics and Cosmology, Ministry of Education, China, and the Science and 
Technology Commission of Shanghai Municipality~(11DZ2260700).
\end{acknowledgments}

\end{document}